\documentclass[journal=jacsat,manuscript=article]{achemso}
\setkeys{acs}{articletitle = true}
\usepackage[version=3]{mhchem} 

\usepackage{amsmath}  
\usepackage{amsfonts} 
\usepackage{subfig}
\usepackage{graphicx} 
\usepackage{booktabs}
\usepackage{threeparttable}
\usepackage[font=footnotesize]{caption}
\usepackage{verbatim}
\usepackage{xcolor}
\usepackage{soul}
\usepackage{float}

\newcommand{\mbf}[1]{\mathbf{#1}}

\author{James Shee}
\email{js4564@columbia.edu}
\author{Benjamin Rudshteyn}
\affiliation{Department of Chemistry, Columbia University, 3000 Broadway, New York, NY, 10027}
\author{Evan J. Arthur}
\affiliation{Schrödinger Inc., 120 West 45th Street, New York, NY 10036}
\author{Shiwei Zhang}
\affiliation{Center for Computational Quantum Physics, Flatiron Institute, 162 5th Avenue, New York, NY 10010}
\alsoaffiliation{Department of Physics, College of William and Mary, Williamsburg, VA 23187}

\author{David R. Reichman}
\author{Richard A. Friesner}
\affiliation{Department of Chemistry, Columbia University, 3000 Broadway, New York, NY, 10027}

\title{\large On Achieving High Accuracy in Quantum Chemical Calculations  of 3$d$ Transition Metal-containing Systems: A Comparison of Auxiliary-Field Quantum Monte Carlo with Coupled Cluster, Density Functional Theory, and Experiment for Diatomic Molecules}

\begin{document}

%
%
%

\begin{abstract}

The bond dissociation energies of a set of 44 3$d$ transition metal-containing diatomics are computed with phaseless auxiliary-field quantum Monte Carlo (ph-AFQMC) utilizing a correlated sampling technique.  We investigate molecules with H, N, O, F, Cl, and S ligands, including those in the 3dMLBE20 database first compiled by Truhlar and co-workers with calculated and experimental values that have since been revised by various groups.  In order to make a direct comparison of the accuracy of our ph-AFQMC calculations with previously published results from 10 DFT functionals, CCSD(T), and icMR-CCSD(T), we establish an objective selection protocol which utilizes the most recent experimental results except for a few cases with well-specified discrepancies.  With the remaining set of 41 molecules, we find that ph-AFQMC gives robust agreement with experiment superior to that of all other methods, with a mean absolute error (MAE) of 1.4(4) kcal/mol and maximum error of 3(3) kcal/mol (parenthesis account for reported experimental uncertainties and the statistical errors of our ph-AFQMC calculations).  In comparison, CCSD(T) and B97, the best performing DFT functional considered here, have MAEs of 2.8 and 3.7 kcal/mol, respectively, and maximum errors in excess of 17 kcal/mol for both methods.  While a larger and more diverse data set would be required to demonstrate that ph-AFQMC is truly a benchmark method for transition metal systems, our results indicate that the method has tremendous potential, exhibiting unprecedented consistency and accuracy compared to other approximate quantum chemical approaches.
\end{abstract}


\section{Introduction}
Transition metals play a vital role in a wide range of important processes in biology \cite{trautwein1997bioinorganic} and materials science \cite{khomskii2014transition}. Many redox and catalytic reactions, such as the water splitting reaction in Photosystem II,\cite{askerka2016oec} are dependent upon the electronic structure of specific transition metal-containing clusters.  A precise understanding of the chemistry and physics of these processes at an atomic level of detail can only be elucidated by accurate quantum chemical calculations in conjunction with extensive experimental data. However, quantum chemical methods have had great difficulty in the treatment of transition metal-containing systems.\cite{harvey2006accuracy,friesner2017localized} Even for small molecules, the accuracy of high level \textit{ab initio} approaches for these systems has been far from clear. For larger systems, density functional theory (DFT) has been the only viable alternative.  Much has been learned from  applying DFT to complex systems,\cite{ziegler1991approximate} but while in many cases surprisingly good quantitative results have been obtained, there are also cases where errors as large as 40 kcal/mol can be observed.\cite{aoto2017arrive} A benchmark quality quantum chemical methodology which can be scaled up efficiently to treat systems 30-100 atoms in size would be a transformative advance.

Validation of benchmark accuracy must start with molecules containing only a few atoms, as was the case for organic systems, where coupled cluster (CC) based approaches, predominantly CCSD(T), have been able to demonstrate accuracy to better than 1 kcal/mol, with steady, systematic improvement over the past 20 years \cite{bartlett2007coupled}. For transition metals, the challenge is compounded by uncertainties in many of the experimental measurements used as relevant test cases, as is apparent in recent investigations using a variety of computational methods on small molecules.\cite{bross2013explicitly,manivasagam2015pseudopotential,jiang2012comparative,zhang2013tests,moltved2018chemical,jiang2011toward,carlson2014multiconfiguration,bao2017predicting,baobimetallic,sharkas2017multiconfiguration,tran2018spin}
Focusing on CC methods, electronic excitations for atoms are well described by CCSD(T) calculations using large basis sets and correcting for relativistic effects \cite{bartlett2007coupled}. However, even for problems involving simple diatomic molecules, such as the dissociation energy of NiH, there is considerable uncertainty as to the degree of accuracy that CCSD(T) methods can achieve \cite{aoto2017arrive}. Error bars in the experimental gas phase measurements of dissociation energies of transition metal-containing diatomics are reported to be as large as $\sim$ 5-10 kcal/mol in unfavorable cases (and for a few experiments may exceed that threshold) \cite{aoto2017arrive}.  With this level of possible error, it is very challenging to carry out robust statistical assessments of various approaches, as was done successfully for organic systems using the G2 \cite{curtiss1998assessment} and G3 \cite{curtiss2000assessment} databases of Pople and co-workers. 

Over the past decade, there have been a number of efforts to evaluate the accuracy of CC approaches for small transition metal-containing molecules.  The most recent work over the past 5 years has focused principally on diatomic species.  The electronic structure problem is still qualitatively more difficult than it is for atoms, but the minimal size of the system enables very high level theoretical methods to be applied on relatively large data sets, and the experimental errors are in general more well controlled than for more diverse test cases (although severe individual problematic cases remain).  In addition to the experimental uncertainty, a key issue that has emerged is that the CC numbers can vary considerably depending upon the details of the calculations. The treatment of relativistic effects, spin-orbit coupling, and basis set extrapolation can have large effects on the accuracy of predicted bond dissociation energies.  Early work from this period did not necessarily utilize a complete treatment of such aspects. For example, Ref. \citenum{truhlar} employed single point calculations only in the triple zeta basis set, without any basis set extrapolation. Subsequent work has established standard protocols (which we discuss in more detail below) which appear to be sufficient to handle these particular aspects of the problem to near-chemical accuracy.\cite{fang2017prediction, cheng2017bond}  Nevertheless, significant discrepancies between theory and experiment remain, and have been challenging to analyze definitively. 

The current state of the art is well reflected in the recent work of de Oliviera-Filho and coworkers. They consider the bond dissociation energies of 60 diatomic species, each consisting of one transition metal atom and one hydrogen or second or third row acceptor.  Of these systems, 42 contain a first row transition metal, to which we will limit our consideration in the present work (we plan to consider higher row transition metals in subsequent work). This data set of diatomics is expanded in size as compared to earlier efforts along the same lines, e.g. the 3dMLBE20 data set of Truhlar and coworkers, which contains 20 molecules, 19 of which are included in Ref. \citenum{truhlar}.  All of the test cases have available experimental results that are at least plausible, although the issues with uncertainy noted above remain. We adopt the data set of de Oliveira-Filho and co-workers as a starting point for our analysis in what follows, adding and subtracting a few cases based on consideration of the experimental results, as will be discussed in detail below.  A larger and more diverse data set enables more robust conclusions to be drawn concerning the performance of quantum chemical approaches in thermochemical calculations. Calculated errors can vary dramatically among molecules that are apparently very similar, as can be seen by examining the performance of DFT methods in calculating atomization energies for molecules in the G3 database (222 molecules).\cite{friesner2006localized}  While the present data set is in our view not sufficiently large or diverse to draw rigorous conclusions concerning benchmark quality (on the order of 1 kcal/mole mean absolute error (MAE) across the entire range of first row transition metal chemistry), it does represent a reasonable place to start an assessment of whether a given method is a candidate for such performance, assuming that the experimental errors can be sufficiently well understood.  

CC-based calculations are carried out in reference \citenum{aoto2017arrive} at the state of the art level, carefully converging results to the complete basis set (CBS) limit and incorporating core-valence and relativistic effects. In addition to single-reference (SR) CCSD(T) calculations, multi-reference (MR) CCSD(T) calculations are also reported.  Such computations require nontrivial approximations, due to the potentially large computational expense incurred by the use of multi-reference wavefunctions. Nevertheless, it is of great interest to observe the effects of attempting to employ a methodology that, in principle, represents a systematic improvement over CCSD(T), addressing the well known presence of multiple relevant low-lying states in the electronic structure of transition metals.  The results presented in that work provide a qualitative picture of the accuracy of CC based approaches for transition metal-containing systems. In many cases, both the SR and MR approaches are within a few kcal/mole of the experimental value of the dissociation energy.  In others, the MR calculation provides a dramatic correction to SR results that were in considerable disagreement with experiment, by as much as 14.6 kcal/mole. In still other cases, the MR results continue to exhibit large disagreements with experiment, up to 11.6 kcal/mole. For these remaining outliers, even at the best (MR-CCSD(T)) level of theory employed, the question remains as to the relative contribution of computational and experimental errors to the discrepancies. A reasonable conclusion to be drawn is that SR-CCSD(T) is not capable of benchmark quality results for transition metal-containing systems (in contrast to non-metal systems, where MAEs $<$ 1 kcal/mole have been reported for a subset of the G2 database\cite{feller2001extended}).

An alternative approach to CC calculations that is, in principle, capable of achieving systematically improvable and benchmark-quality accuracy for transition metal-containing systems is the phaseless auxiliary-field quantum Monte Carlo (ph-AFQMC) methodology \cite{zhang1995constrained,zhang2003quantum,al2006auxiliary,motta2018ab,zhang2018ab}.
ph-AFQMC is a stochastic method capable of predicting observables of chemical systems with high accuracy, and has been used to benchmark a variety of strongly correlated electronic systems,\cite{al2007bond,motta2017Hchain,zheng2017stripe} including transition metal-containing species.\cite{al2006oxides,virgus2012ab,purwanto2013frozen,virgus2014stability,purwanto2015auxiliary,ma2015quantum,purwanto2016auxiliary,ma2017auxiliary,shee2018gpu,moralesNiO}  
A feature of the method is that its computational cost scales with the fourth power of the system size (cubic scaling has been demonstrated for larger systems\cite{motta2018efficient}), but to date ph-AFQMC has not been widely applied to molecular systems.  A few calculations have been done on larger systems, but these required large amounts of computational power due to the presence of a large prefactor.\cite{motta2018ab,zhang2018ab}

In a recent series of papers, we have described a number of technical advances which have demonstrated dramatic reductions in the computational requirements for ph-AFQMC calculations, while in some cases actually improving their accuracy and robustness.  The first of these is the use of correlated sampling\cite{shee2017chemical}.  With correlated sampling, energy differences between two states are computed by sampling both states with the same set of auxiliary fields, leading to significant cancellation of error.  This enables energy differences to be computed in a much shorter amount of propagation time and with fewer samples than would normally be required to obtain a given statistical error\cite{shee2017chemical}. Furthermore, these measurements at short propagation times are often converged before the full accumulation of the errors associated with the phaseless approximation, thus yielding results that are closer to the unbiased, exact value\cite{shee2017chemical}. The second advance is the development of an efficient implementation of ph-AFQMC on graphical processing units (GPUs), including the use of the Sherman-Morrison-Woodbury (SMW) algorithm to accelerate calculations using multideterminental trial wavefunctions\cite{shee2018gpu}. For problems where correlated sampling is applicable, the combination of these two techniques can reduce the computational effort by more than two orders of magnitude, enabling the method to be applied to larger systems, and also to substantially larger data sets. Further efficiency improvements are feasible (reducing both the scaling and the prefactor), leading to the possibility that ph-AFQMC will emerge as a scalable benchmark methodology for transition metal-containing systems.

In the present paper, we apply our ph-AFQMC methodology to a subset of the diatomics considered in Ref. \citenum{aoto2017arrive}, specifically all those containing first row transition metals (44 test cases in all).  We have already shown, in Ref. \citenum{shee2018gpu}, that ph-AFQMC yields excellent accuracy for the ionization potentials of first row transition metal atoms. This finding is a good starting point, but it is clear from previous efforts in the literature that diatomic dissociation energies are much harder to compute with kcal/mol accuracy\cite{aoto2017arrive}, and that the validation problem is more challenging given the issues with the experimental data. 

The first objective of this paper is to address key methodological issues that are critical to achieving robust and accurate results with ph-AFQMC for diatomic dissociation energies. Firstly, we demonstrate that correlated sampling can be made to work well for heavy atom dissociation, building on previous work which only considered removal of a hydrogen atom\cite{shee2017chemical}. We find that correlated sampling not only provides substantial reductions in computational effort, but is essential in obtaining accurate energetics for these systems. The ability to treat heavy atom dissociation substantially expands the domain of applicability of correlated sampling to a wide range of chemical and biological problems. 

Secondly, in ph-AFQMC calculations it is essential to utilize a sufficiently ``good'' trial function.  We explore CASSCF type wavefunctions\cite{roos2007complete} for this purpose,  and take advantage of the fact that for these small systems the dissociation energies can be converged with respect to active space size.  The ph-AFQMC calculations for the diatomic molecules in our test set used between 100 and 5700 determinants. Our efficient GPU implementation of the SMW approach is essential for the utilization of large multideterminant trial functions of this form while keeping the increase in computer time at only a small factor. 

Thirdly, we investigate three different approaches to estimating the CBS limit.  All strategies employ a ph-AFQMC calculation in the triple zeta basis, and two-point extrapolations based on second-order M\o ller-Plesset Perturbation Theory (MP2)\cite{moller1934note}, CCSD(T), and entirely based on ph-AFQMC. MP2 extrapolation suffices for many, but not all cases. CCSD(T) extrapolation usually does better, if not similarly to MP2. For a subset of the cases which we found to be exceptionally difficult, we show that direct AFQMC extrapolation is consistently able to improve the MP2 and CCSD(T) results.

Fourthly, we include new experimental values published in Ref. \citenum{morse2018predissociation}, and we identify one case (ZnS) where we believe that the experimental result is problematic, i.e. outside the error bars reported in the experimental papers.  The very large discrepancies of experiment with both state-of-the-art CC and QMC results, along with a detailed analysis of the experiments, lead us to believe that the experiment is in error.  Theory cannot evolve to benchmark status without such conclusions being drawn along the way.  With an optimized methodology defined, and with an objectively chosen set of reference values, we find remarkably good agreement between the ph-AFQMC results and the experimental data (taking into account the experimental error bars).  We compare the MR-CCSD(T) and CCSD(T) values reported in Ref. \citenum{aoto2017arrive} to the reference values, and find that the CC methods display a number of large outliers (fewer for the MR-corrected version).  We also analyze the performance of 10 DFT functionals, reported in Ref. \citenum{aoto2017arrive}.  Assessment of DFT results has been a feature of many of the papers cited above; however, the accuracy of the assessment has been problematic due to the uncertain nature of the reference values.  

Finally, we discuss computational efficiency and the feasibility of scaling up to larger systems. It is possible to parallelize AFQMC efficiently across a large farm of GPUs (we plan to report the results of such an implementation in the near future), so with sufficient computational hardware resources, AFQMC calculations with a large number of basis functions can be carried out in a reasonable wall clock time. Furthermore, significant improvements in the AFQMC algorithm are still possible, and likely will be necessary to handle grand challenge problems with the goal of achieving true benchmark status. As noted above, the generation of sufficiently good trial functions may turn out to be the leading challenge to be faced in this scale up effort. 

Our paper is organized as follows.  In Sec. II, we detail the extension of our correlated sampling approach to the computation of bond dissociation energies.  In Sec. III, we provide additional computational details, and Sec. IV includes a discussion of the landscape of experimental methods.  In Sec. V we present our results for the $D_e$'s of the 3$d$ transition metal diatomics, and justify our selection of the reference values used in the comparative statistical analysis of the various computational methods.  In Sec. VI, we offer concluding remarks.

\section{Correlated Sampling for BDEs}
Recently we have introduced a correlated sampling (CS) approach for quantities involving energy differences which is capable of reducing computational prefactors\cite{shee2017chemical} and in some cases the severity of the phaseless approximation.\cite{shee2018gpu}  In this section, we show that significant reductions in statistical errors are obtained not only for hydrogen abstraction reactions, as shown previously, but also for bond breaking events between a transition metal and a heavier ligand atom.

For diatomic molecules consisting of a metal (M) and ligand (L), the following equation for the bond dissociation energy is employed:
\begin{equation}
D_e = E(M) + E(L) - E(ML) + \Delta SO.
\label{De}
\end{equation}
We use CS to compute $E(M) - E(ML)$, where in the former term so-called ``ghost'' basis functions centered at the coordinates of $L$ are added, but without the nuclear charge or electrons from the ligand.  We note that the basis set superposition imbalance\cite{simon1996does}, if any, that is introduced at the TZ level vanishes in the CBS limit.  $E(L)$ is computed using the population control (PC) method detailed in Refs. \citenum{shee2017chemical} and \citenum{nguyen2014cpmc}, and $\Delta SO$ is the calculated energy difference due to spin-orbit coupling taken from Ref. \citenum{aoto2017arrive}. 

\begin{figure}[h]
    \centering
    \includegraphics[width=13cm]{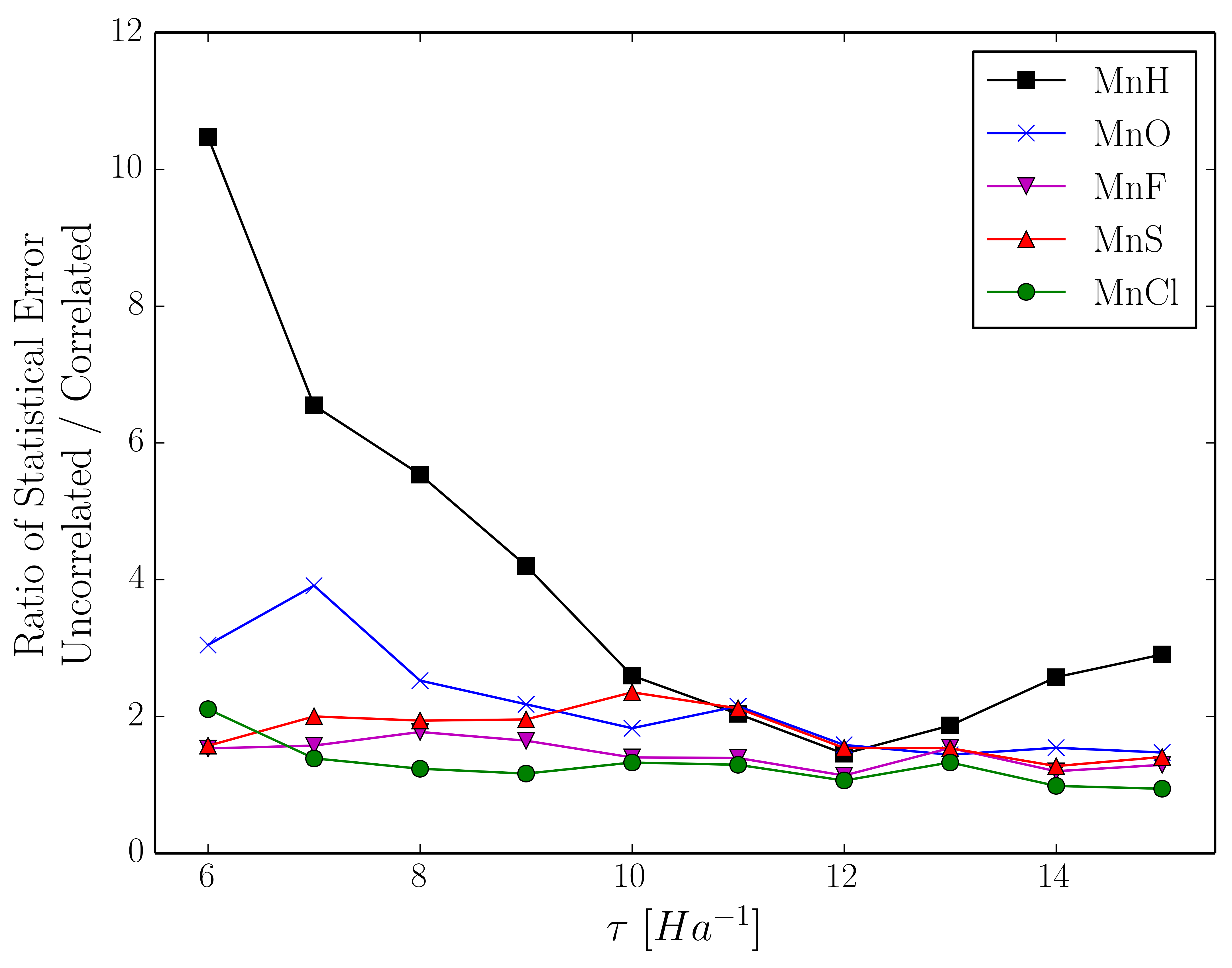} 
    \caption{Ratio of the standard errors, as a function of imaginary time, resulting from ph-AFQMC calculations with correlated vs uncorrelated sampling, for the five Mn-containing diatomic species. } 
    \label{fig:CorrUncorr}
\end{figure}

The reduction in statistical error, compared to the uncorrelated sampling approach, is shown in Fig. \ref{fig:CorrUncorr} for the Mn-containing diatomics in our set (the saving in computational time is, as usual, given by the square of the error ratio).  We find that the effect of correlated sampling is largest for the hydride ligand, with decreasing noise reduction efficiency as the ligand atomic number increases.  

In Ref. \citenum{shee2018gpu}, CS results exhibit equivalent or \emph{better} accuracy compared to the conventional method of running ph-AFQMC calculations with PC employing the same trial wavefunctions, for the ionization potentials of first row transition metal atoms.  
We find the same behavior in the calculation of the BDEs in this work.  For ph-AFQMC calculations of the $D_0$ of MnCl in the aug-cc-pwCVTZ-DKH basis, PC and CS yield values of 86(1) and 82(2) kcal/mol, compared with the experimental value of 80(2).  For MnS PC, CS, and experimental $D_0$ values are 78(1), 71(1), and 70(3).  Thus, in light of significantly improved computational efficiency \emph{and} accuracy, we use CS for the $E(M) - E(ML)$ part of all the BDE calculations in this work.

\section{Computational Details} 

We use PySCF\cite{sun2018pyscf} to obtain all inputs required by our ph-AFQMC calculations.  To compute the trial wavefunctions used in this work, we first perform restricted (open-shell) HF calculations, ensuring that the electronic configurations are consistent with the term symbols published in Refs. \citenum{aoto2017arrive} and \citenum{thomas2015accurate}.  Canonical HF orbitals are used to initialize restricted Complete Active Space Self-Consistent Field\cite{roos2007complete} (CASSCF) calculations.  The resulting wavefunctions are truncated such that the sum of the squares of the CI coefficients kept is $> 98\%$ in all cases, resulting in $\sim$800 determinants on average.  

We stress that the spin \emph{and} orbital symmetries, which we enforce at the RHF level, cannot be overlooked.\cite{langhoff1988ab,TiHstates}  For example, the latter can change the computed $D_e$ in TiH by a staggering 15 kcal/mol.  The diatomic term symbols, active space specifications, and bond lengths are shown in Table \ref{table:termSymbols}. 

\begin{table}[H]
\begin{threeparttable}
\caption{Electronic States, Active Spaces, and Bond Distances used in our ph-AFQMC calculations.  X/Y means that both active space configurations produced statistically equivalent results.  The number in parenthesis is the experimental bond length.}
\begin{tabular}{l l l l l}
\hline\hline
                    &  electronic state &  CASSCF Active Space &    $R_e$ [\textup{\AA}] CC (expt)     &     \\ [0.5ex] 
\hline
ScH & $^1\Sigma$ & 10e18o & 1.762 (1.7754) \\
ScO & $^2\Sigma$ & 13e15o & 1.664 (1.6661) \\
ScF & $^1\Sigma$ & 14e15o & 1.787 (1.787) \\
ScS & $^2\Sigma$ & 13e15o & 2.132 (2.1353) \\
TiH & $^4\Phi$ & 10e18o & 1.768 (1.777)\\
TiN & $^2\Sigma$ & 13e15o/7e18o & 1.57 (1.5802)\\
TiO & $^3\Delta$ & 14e15o & 1.617 (1.6203)\\
TiF & $^4\Phi$ & 15e15o & 1.8311 (1.8311)\\
TiS & $^3\Delta$ & 10e18o & 2.0827 (2.0827)\\
TiCl & $^4\Phi$ & 15e15o & 2.2642 (2.2697)\\
VH & $^5\Delta$ & 12e13o & 1.684 (1.730)\\
VN & $^3\Delta$ & 14e15o/10e17o & 1.544 (1.5703)\\
VO & $^4\Sigma$ & 15e15o & 1.5839 (1.5893)\\
VCl & $^5\Delta$ & 16e15o/10e16o & 2.2273 (2.2145)\\
CrH & $^6\Sigma$ & 13e18o & 1.6293 (1.6554)\\
CrO & $^5\Pi$ & 10e16o & 1.6116 (1.615)\\
CrF & $^6\Sigma$ & 11e17o & 1.776 (1.7839)\\
CrCl & $^6\Sigma$ & 17e15o/11e17o & 2.1688 (2.194)\\
MnH & $^7\Sigma$ & 14e18o & 1.727 (1.7309)\\
MnO & $^6\Sigma$ & 17e15o/11e18o & 1.638 (1.6446)\\
MnF & $^7\Sigma$ & 18e15o & 1.834 (1.836)\\
MnS & $^6\Sigma$ & 17e15o/11e18o & 2.0633 (2.0663)\\
MnCl & $^7\Sigma$ & 18e15o/12e18o & 2.2355 (2.2352)\\
FeH & $^4\Delta$ & 9e18o & 1.5478 (1.606)\\
FeO & $^5\Delta$ & 12e17o & 1.612 (1.6164)\\
FeS & $^5\Delta$ & 12e17o & 2.009 (2.0140)\\
FeCl & $^6\Delta$ & 13e17o & 2.1751 (2.1742)\\
CoH & $^3\Phi$ & 10e15o/10e18o & 1.5049 (1.5327)\\
CoO & $^4\Delta$ & 13e17o & 1.6286 (1.5286)\\
CoS & $^4\Delta$ & 13e17o & 1.9786 (1.9786)\\
CoCl & $^3\Phi$ & 14e17o & 2.0749 (2.0656)\\
NiH & $^2\Delta$ & 11e15o/11e19o & 1.4538 (1.4538)\\
NiO & $^3\Sigma$ & 14e17o & 1.626 (1.6271)\\
NiF & $^2\Pi$ & 15e17o & 1.733 (1.7387)\\
NiCl & $^2\Pi$ & 15e17o & 2.0539 (2.0615)\\
CuH & $^1\Sigma$ & 12e15o/12e19o & 1.4593 (1.4626)\\
CuO & $^2\Pi$ & 15e17o & 1.709 (1.7246)\\
CuF & $^1\Sigma$ & 16e17o/10e19o & 1.745 (1.7449)\\
CuS & $^2\Pi$ & 11e19o & 2.051 (2.0499)\\
CuCl & $^1\Sigma$ & 16e17o & 2.0498 (2.0512)\\
ZnH & $^2\Sigma$ & 13e15o & 1.5899 (1.5935)\\
ZnO & $^1\Sigma$ & 16e12o & 1.6989 (1.7047)\\
ZnS & $^1\Sigma$ & 16e17o & 2.0427 (2.0464)\\
ZnCl & $^2\Sigma$ & 17e16o & 2.1274 (2.1300)\\
\hline
\end{tabular}
\label{table:termSymbols}
\end{threeparttable}
\end{table}

Our ph-AFQMC calculations correlate \emph{all} electrons (i.e. no frozen-core), and utilize the ``hybrid" formulation of the algorithm\cite{purwanto2009pressure}.  With an imaginary-time step of 0.005 $Ha^{-1}$, walker orbitals are orthonormalized every other propagation step, and energy measurements are taken every 0.1 $Ha^{-1}$.  
We employ a cutoff of $10^{-4}$ for the Cholesky decomposition of the two-electron integrals.  We have verified that these parameter choices result in biases smaller than the statistical error bar.\cite{shee2018gpu}  We use the aug-cc-pwCVxZ-DKH basis sets\cite{balabanov2005systematically} and the spin-free exact two-component approach\cite{kutzelnigg2005quasirelativistic,peng2012exact} to account for scalar relativistic effects.  For the 3dMLBE20 molecules, the combination of this level of theory and basis sets has produced good results for CC calculations.\cite{cheng2017bond}

The MP2-assisted CBS extrapolation protocol is detailed in Refs. \citenum{shee2017chemical} and \citenum{purwanto2011assessing}.  After a ph-AFQMC calculation in the triplet-zeta (TZ) basis, a CBS correction is obtained by extrapolating the correlation energies as computed with MP2 using the $\frac{1}{x^3}$ form, with $x=3,4$\cite{purwanto2013frozen,helgaker1997basis,balabanov2005systematically}.  We employ a scaling factor, which is the ratio of the MP2 and QMC values in the TZ basis.  For all diatomics we performed both restricted and unrestricted HF calculations to compute correlation energies, and choose the method which leads to a scaling factor closest to 1.  Finally, following Ref.~\citenum{thomas2015accurate} and our own observation that the HF energies converge relatively quickly in this sequence of basis sets, we use the 5Z ($x=5$) value for the CBS HF energies, and add this to the extrapolated correlation energy to arrive at our final result.

For the small molecules considered here, CCSD(T) calculations can be performed in large basis sets, and results have been made available in the Supporting Information of Ref. \citenum{aoto2017arrive}.  To evaluate the reliability of the MP2-assisted protocol for transition metal-containing systems, we use the published CCSD(T) data to extrapolate our ph-AFQMC results to the CBS limit as follows:  We take the difference between de Oliveira-Filho's CCSD(T)(CV)/CBS estimate of $D_e$,  as was obtained via $1/x^3$ extrapolation of the correlation energy at $x$ = Q,5 with respect to the restricted open-shell HF reference, and their value in the aug-cc-pwCVTZ basis.  We then add this term to our ph-AFQMC result in the aug-cc-pwCVTZ-DKH basis.  We estimate the statistical error in the CBS limit using that in the TZ basis combined with their $x$ = T,Q CCSD(T) values.  We note that this procedure assumes that the optimal bond lengths and scalar relativistic contribution to the BDEs are independent of basis size, as is done in Ref. \citenum{aoto2017arrive}, among other works.  The spin-orbit term, $\Delta SO$, in Eq. \eqref{De} is taken from Ref. \citenum{aoto2017arrive}, in which values are computed using CASSCF in a QZ basis.

We emphasize that in both of these extrapolation approaches, AFQMC calculations are only performed in the TZ basis.  In our view, this is a significant source of computational expedience, as the convergence of observables with the size of the CASSCF active space used in the trial function is expected to be slower in basis sets of increasing size.  For large chemical systems with substantial multireference character we note that other methods such as CASPT2 or even ph-AFQMC with a single-determinant trial wavefunction can be used to compute a CBS correction.

For a select number of cases where we encountered significant discrepancies among our calculated methods and with respect to experiment, we perform ph-AFQMC calculations in both the TZ and QZ basis sets, and extrapolate to the CBS limit.  We view this extrapolation protocol to be of the highest quality, and for the purposes of this paper we employ it as required.  

\begin{figure}[H]
    \centering
    \includegraphics[width=13cm]{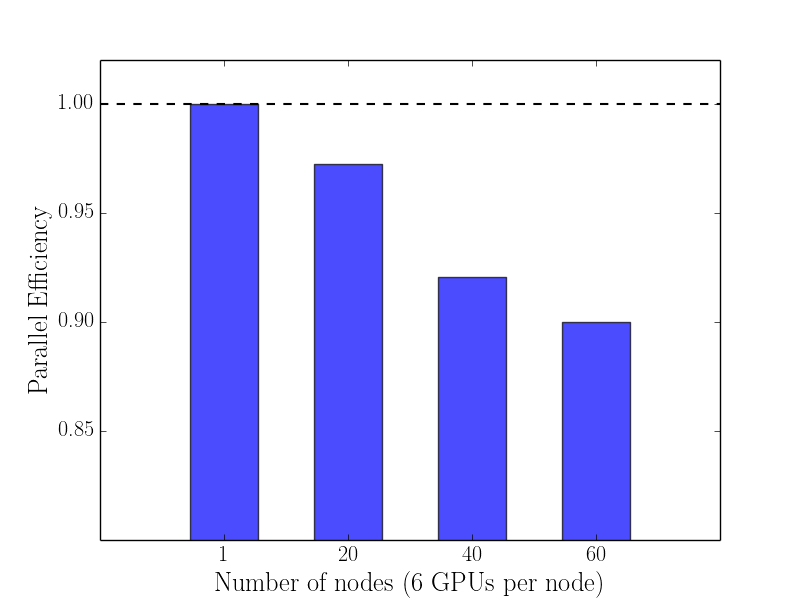} 
    \caption{For a set of node counts on the Summit computing cluster, we plot the parallel efficiency, defined as the speed-up over a 1 node calculation divided by the number of nodes.  Each node utilizes 6 NVIDIA V100 GPU cards. } 
    \label{fig:GPU_PE}
\end{figure}

All ph-AFQMC calculations use single precision floating point arithmetic (which we have verified to give consistent results within statistics as double precision calculations\cite{shee2018gpu}) and were run on NVIDIA GeForce GTX 1080, Tesla P100 and V100 graphical processing units.  Our code is parallelized with Message Passing Interface (MPI), and we observe excellent strong-scaling parallel efficiency, shown in Fig. \ref{fig:GPU_PE} for the CoO diatomic in the QZ basis.  Using 360 GPUs on 60 nodes of the Summit supercomputer, the parallel efficiency of our implementation is still 90$\%$.  This allows us to run large calculations in minutes, a capability not possible for traditional, non-stochastic quantum chemical methods. 

Not all of the ``experimental'' bond lengths in Ref. \citenum{truhlar} were actually obtained from experiments, so we use calculated bond lengths for the 3dMLBE20 set at the CCSD(T)/CBS level of theory, taken from Ref. \citenum{fang2017prediction}.  When the experimental bond lengths, as given in Ref. \citenum{aoto2017arrive}, differ by more than 0.01\AA \ from the CCSD(T) values, we ran ph-AFQMC calculations at both bond lengths. In the future we will consider using geometries optimized within ph-AFQMC.\cite{motta2018communication}

The active spaces utilized to generate the trial wavefunctions were initially chosen with the intention of realizing a maximal cancellation in the systematic error associated with finite active spaces.  That is, the number of active electrons (orbitals) of the isolated metal and ligand should sum to the number of electrons (orbitals) in the metal-ligand dimer.  However, for these diatomic systems we found it possible to converge the BDE with respect to increasing active space sizes, and prioritized this convergence at times over the balanced protocol described above.  The size of the active spaces is limited by the current CI module in PySCF, yet we were able to employ active spaces with up to 19 orbitals, allowing for satisfactory convergence throughout.  

For the isolated ligands, we confirmed the convergence of the energy from ph-AFQMC/PC calculations with increasingly large active space sizes, and found that in all cases except for the sulfur and fluorine atoms, CASSCF did not lower the energy by more than a milliHartree with respect to unrestricted Hartree Fock (UHF).  Hence, we use UHF for these cases, CASSCF(6e,8o) for S, and CASSCF(7e,16o) for F.  The latter is consistent with our previous work,\cite{shee2017chemical} in which we found that an active space of this size was necessary to obtain chemically accurate electron affinities for the F atom.  

\section{Experimental BDEs}

The experimental BDEs given by Truhlar and co-workers for the 3dMLBE20 set were determined either from experimental enthalpies of formation (TiCl, VH, VO, VCl, CrO, CrCl, MnS, MnCl, FeCl, CoCl, NiCl, CuCl, ZnH, ZnO, ZnS, and ZnCl), which have their own error bars, or from direct measurements of $D_0$ at 0 K (CrO, FeH, CoH, and CuH).\cite{truhlar} Both were converted to $D_e$ via scaled DFT calculations of zero-point energies.

In the follow-up work by Dixon and co-workers, the experimental $D_e$'s for the hydride diatomics were replaced by values derived from hydride transfer experiments.\cite{fang2017prediction} These experiments involve the following reaction

\begin{equation}
M^+ + RH \rightarrow MH + R^+.
\label{hydride}
\end{equation}

By combining the energy of this reaction (referred to in Ref. \citenum{fang2017prediction} as $E_{threshold}$) with the ionization potential (IP) of the metal ($M$), 
the electron affinity (EA) of the hydrogen atom,
and the heterolytic bond dissociation energy of a C-H bond in an organic molecule (typically a hydrocarbon or amine) (BDE$_{heterolytic}$(R-H))
the BDE of the metal hydride is obtained:
\begin{equation}
BDE(MH) = BDE_{heterolytic}(R-H) - IP(M) - E_{threshold} - EA(H).
\label{MHBDE}
\end{equation}

In these measurements, IP($M$), EA($H$), and $E_{threshold}$ are known relatively accurately.
BDE$_{heterolytic}$(R-H) values have more uncertainty, but Dixon and co-workers confirmed the experimental quantities with G3MP2 calculations.\cite{fang2017prediction}  However, the quantity BDE$_{heterolytic}$(R-H) - E$_{threshold}$ is not constant for various R, often varying up to 10 kcal/mol. Therefore, Dixon and co-workers give two values: one that is the average of all the measurements with different R's and one measurement that is the closest to their calculated CCSD(T)-level value. \cite{fang2017prediction}  In the present work, when referring to the values from Dixon and co-workers we only consider the measurements derived from the former method (averaged values).   Stanton and co-workers use similar experimental values for VH and CrH using hydride transfer reactions and also confirm the validity of the BDE$_{heterolytic}$(R-H) using their own HEAT345-Q protocol.\cite{cheng2017bond} These experimental values for the $D_e$ may be an underestimate of the true $D_e$ as the $E_{threshold}$ may be affected by competition with side reactions, which may explain some of persistent disagreement between theory and experiment.\cite{cheng2017bond}

Dixon and co-workers also replaced the values for the chlorides with direct mass spectrometric measurements using Ag-M-Cl vapors, the value for VO with a direct measurement using a Eu-V-O system, and the value for ZnO with a mass spectrometry experiment, and the value for other compounds, particularly ZnS using different, fully experimental, heats of formation using more accurate Joint Army-Navy-NASA-Air Force (JANAF) thermochemical values.\cite{fang2017prediction}  Their selection of best experimental values for the 3dMLBE20 set are listed in Table 5 of Ref. \cite{fang2017prediction}.

Recently, Morse has reviewed his group's progress in obtaining highly precise measurements using resonant two-photon ionization spectroscopy to obtain predissociation thresholds that are equivalent to the BDE's of those diatomics with a very high density of states.\cite{morse2018predissociation} This experiment works by increasing the frequency of the incoming laser pulse until the excited state cation can no longer be detected (the predissociation threshold), because it has dissociated from the excited state's rovibrational state to the ground-state separated atom limit via other unstable excited states. Thus, this technique requires there to be a high density of states to ensure the method is precise and accurate, which precludes study of diatomics containing Cr, Mn, Cu, or Zn. For molecules where this technique is appropriate, it is more precise than many high-temperature Knudsen effusion measurements of gas-phase equilibria  and guided ion beam mass spectrometry. Morse also shows that the measurements are also amenable to testing via a thermodynamic cycle with other precise measurements.\cite{morse2018predissociation}  In our study we convert $D_0$ to $D_e$ using the ZPE data in Ref. \citenum{aoto2017arrive}.

In the study by de Oliveira-Filho and co-workers only spectroscopically-derived data is referenced,\cite{aoto2017arrive} which may explain their omission of the VH molecule.

\section{BDEs of Transition metal-containing Diatomics}

In this section, we show our computed values of $D_e$ for the set of diatomic molecules containing first row transition metal atoms, and compare the calculated ph-AFQMC results to experiments, and to the MR-CCSD(T) calculations performed in Ref. \citenum{aoto2017arrive}. 

As we expect the finite basis set error to be less sensitive to the method used to calculate the correlation energy, we examine various strategies to minimize the compute time required to reach the CBS limit.  The MP2-assisted and CC protocols produced CBS results that are very similar in the majority of cases. Importantly, an extreme value (with respect to unity) of the scaling factor, which is the ratio of the correlation energies at the MP2 and ph-AFQMC levels in the TZ basis, can serve to flag an unreliable MP2 extrapolation.  For instance, the scaling factors for CrH and NiH are both larger than 2.5.  The resulting MP2-assisted predictions for $D_e$ in the CBS limit for NiH was the furthest from experiment that we observed.

The CC method of CBS extrapolation is more reliable, and we checked that all scaling factors are between 0.8 and 1.2.  When the resulting CBS BDE value still differed substantially from experiment, i.e. for ScH, TiS, CrO, CrF, CrCl, CoH, NiH, NiO, NiCl, CuO, and ZnS, we performed ph-AFQMC calculations in both the TZ and QZ basis sets, and extrapolated to the CBS limit.  In all of these cases, except for CrO and ZnS, this procedure produced results consistent (within experimental uncertainty and statistical error) with experimental values.  These two cases will be discussed in detail below.  As control cases, we utilized this procedure for two cases, CoO and CrO, for which the CC CBS estimate is already accurate.  As expected, the pure ph-AFQMC CBS results produced essentially the same values.  

For all species in which the CC and experimental bond lengths, shown in Table \ref{table:termSymbols}, differed by more than 0.1 \textup{\AA}, we performed ph-AFQMC calculations at both bond lengths.  No significant differences in $D_e$ resulted, so the results corresponding to the CC values are shown.  

\begin{figure}[H]
    \centering
    \includegraphics[width=13cm]{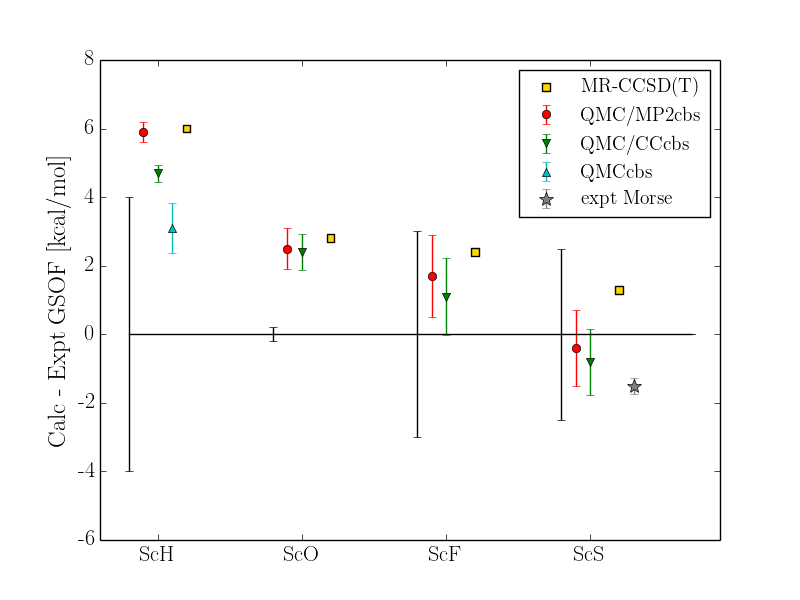} 
    \caption{Deviations [kcal/mol] of various calculations and alternate experiments (when relevant) from the experimental values used by de Oliveira-Filho (GSOF) and co-workers in Ref. \citenum{aoto2017arrive}.  For calculations and experiments, error bars represent statistical error and quoted experimental uncertainties, respectively.} 
    \label{fig:Sc}
\end{figure}

\begin{figure}[H]
    \centering
    \includegraphics[width=13cm]{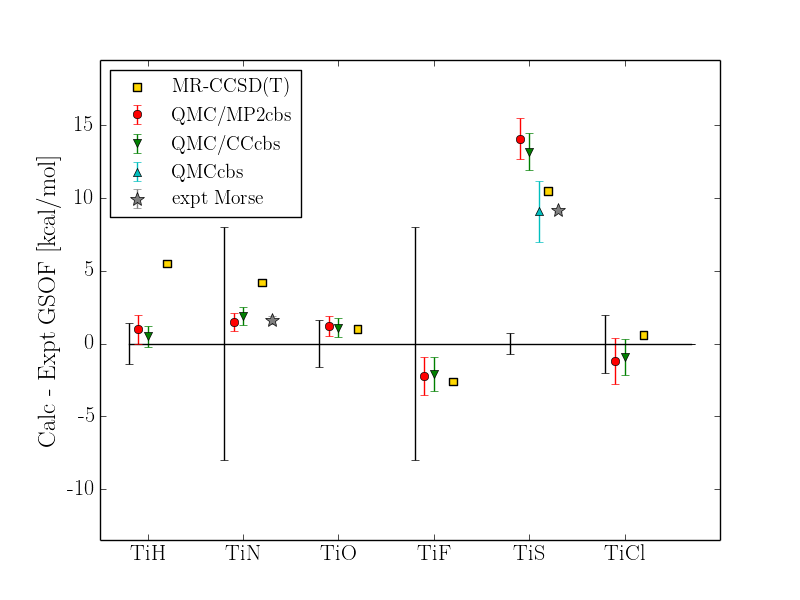} 
    \caption{Same as Fig. \ref{fig:Sc}, but for Ti-containing diatomics.} 
    \label{fig:Ti}
\end{figure}

\begin{figure}[H]
    \centering
    \includegraphics[width=13cm]{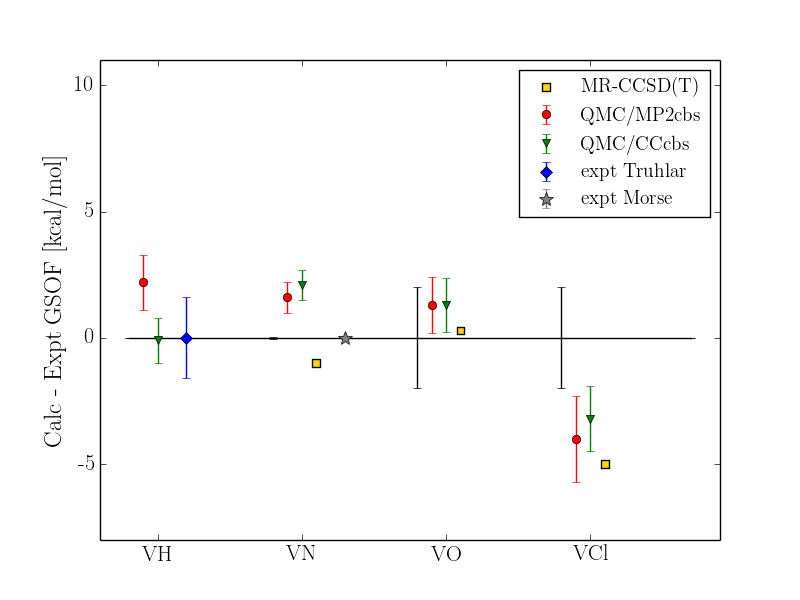} 
    \caption{Same as Fig. \ref{fig:Sc}, but for V-containing diatomics.  VH was not considered in Ref. \citenum{aoto2017arrive}, and we show the experimental result selected by Truhlar and co-workers in Ref. \citenum{truhlar}.} 
    \label{fig:V}
\end{figure}

\begin{figure}[H]
    \centering
    \includegraphics[width=13cm]{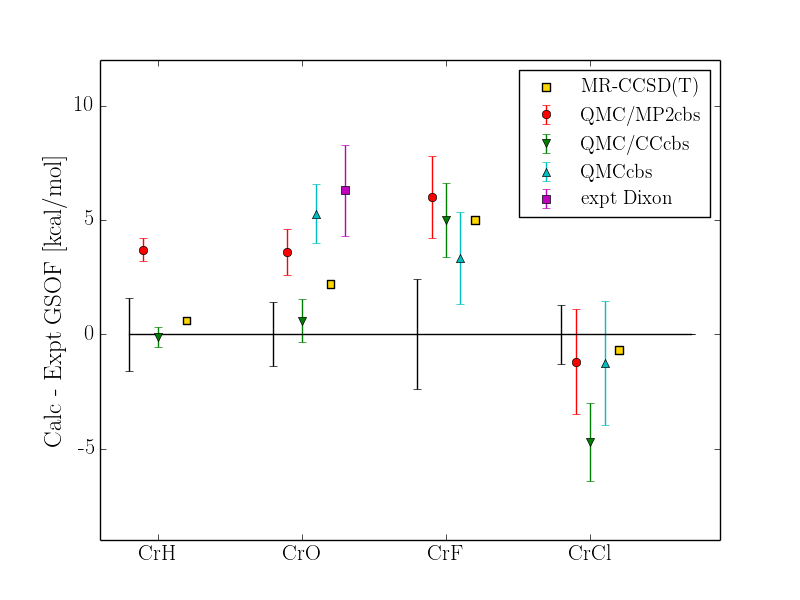} 
    \caption{Same as Fig. \ref{fig:Sc}, but for Cr-containing diatomics.  For CrO we also show the experiment selected by Dixon and co-workers in Ref. \citenum{fang2017prediction}, since it is not consistent with that chosen by de Oliveira-Filho and co-workers\cite{aoto2017arrive}, given the reported uncertainties.}
    \label{fig:Cr}
\end{figure}

\begin{figure}[H]
    \centering
    \includegraphics[width=13cm]{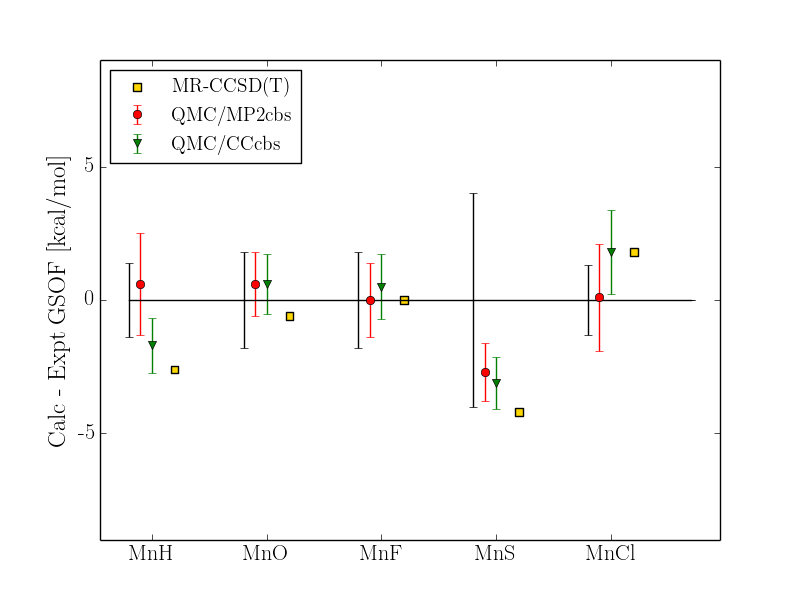} 
    \caption{Same as Fig. \ref{fig:Sc}, but for Mn-containing diatomics.} 
    \label{fig:Mn}
\end{figure}

\begin{figure}[H]
    \centering
    \includegraphics[width=13cm]{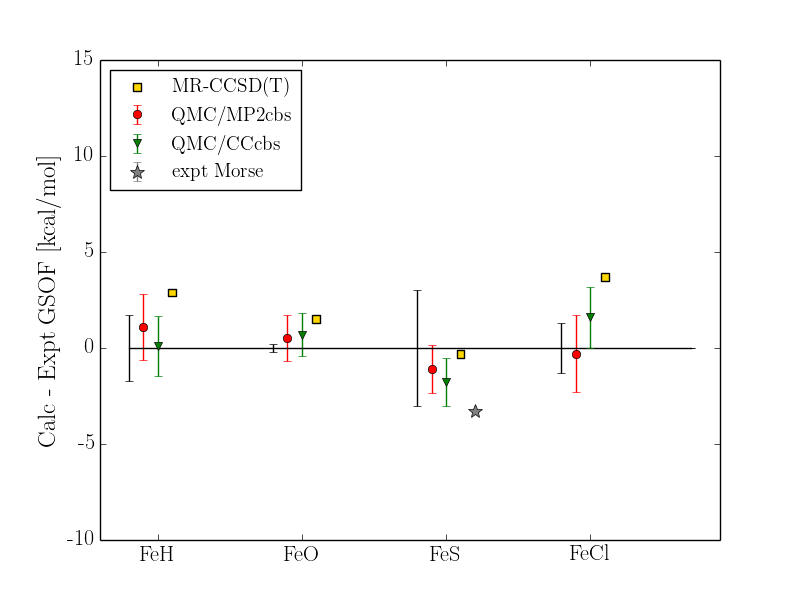} 
    \caption{Same as Fig. \ref{fig:Sc}, but for Fe-containing diatomics.} 
    \label{fig:Fe}
\end{figure}

\begin{figure}[H]
    \centering
    \includegraphics[width=13cm]{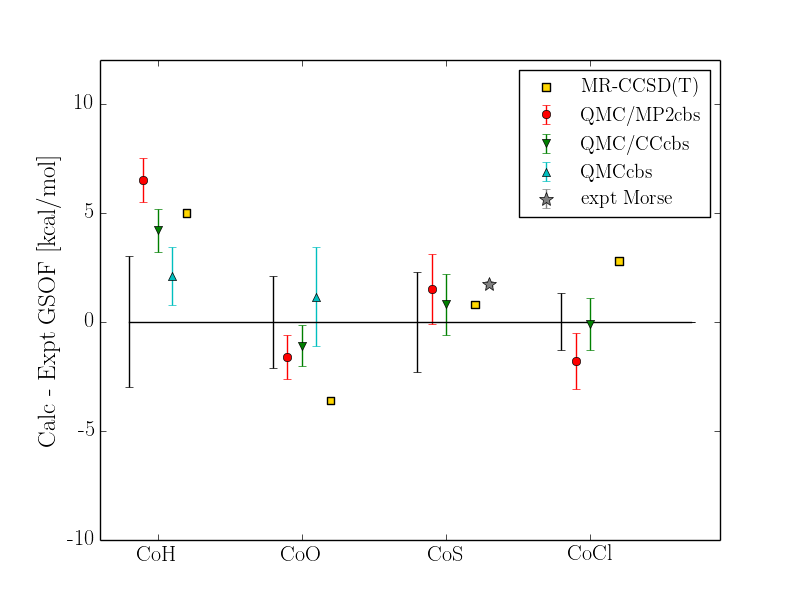} 
    \caption{Same as Fig. \ref{fig:Sc}, but for Co-containing diatomics.} 
    \label{fig:Co}
\end{figure}

\begin{figure}[H]
    \centering
    \includegraphics[width=13cm]{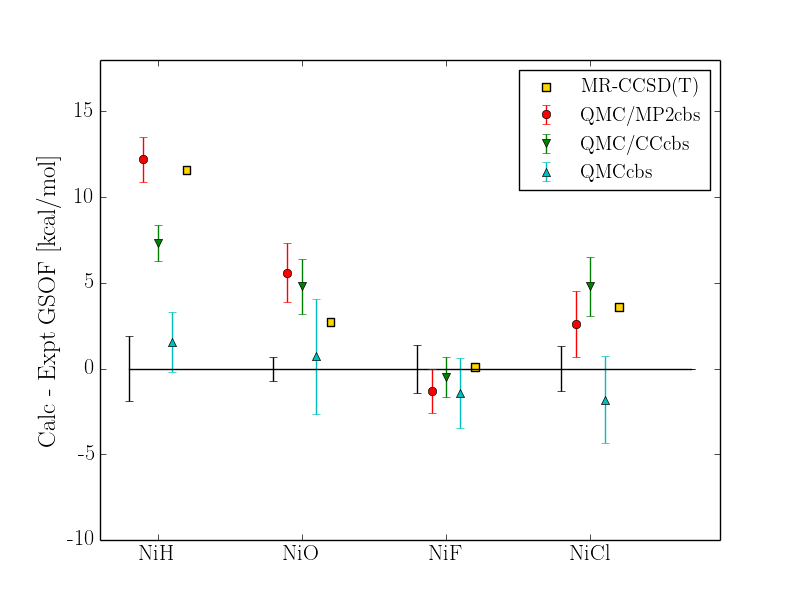} 
    \caption{Same as Fig. \ref{fig:Sc}, but for Ni-containing diatomics.} 
    \label{fig:Ni}
\end{figure}

\begin{figure}[H]
    \centering
    \includegraphics[width=13cm]{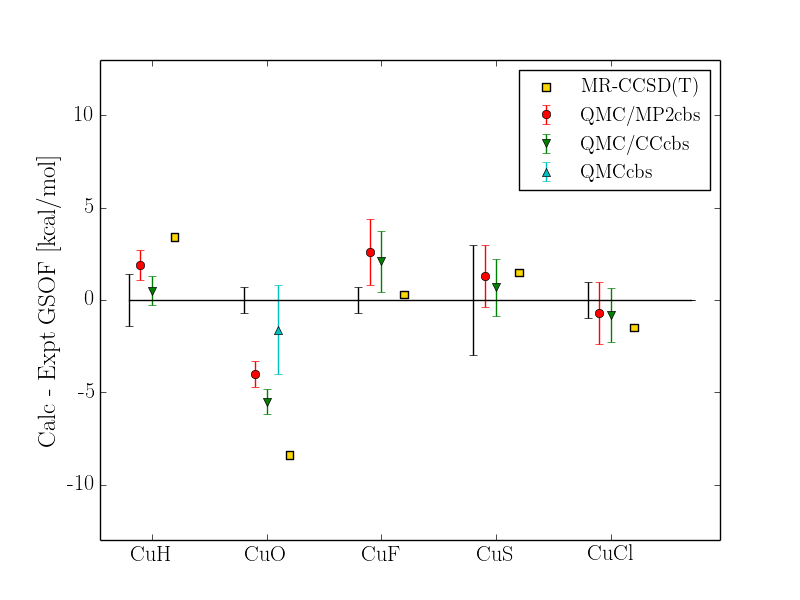} 
    \caption{Same as Fig. \ref{fig:Sc}, but for Cu-containing diatomics.}
    \label{fig:Cu}
\end{figure}

\begin{figure}[H]
    \centering
    \includegraphics[width=13cm]{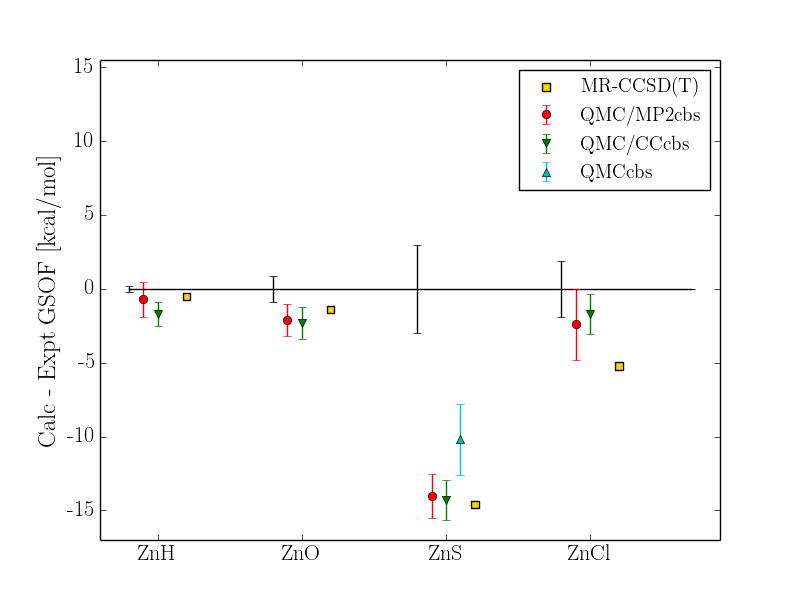} 
    \caption{Same as Fig. \ref{fig:Sc}, but for Zn-containing diatomics.} 
    \label{fig:Zn}
\end{figure}

In Figs. \ref{fig:Sc} - \ref{fig:Zn} we present our results by metal species.  We show the selection of experimental values and the MR-CCSD(T) results from Ref. \citenum{aoto2017arrive} by default.  We also show experimental values from the predissociation technique of Ref. \citenum{morse2018predissociation} when available, and note their extremely small uncertainties.  We considered the experimental selections in Ref. \citenum{fang2017prediction} as well, and plot their choice only when it is not within error bars of the corresponding value from Ref. \citenum{aoto2017arrive}.   

We would like to highlight that the measurements in Ref. \citenum{morse2018predissociation} were published \emph{after} the ph-AFQMC calculations in this work were completed.  It is rather remarkable that in all six relevant cases - ScS, TiN, TiS, VN, and FeS - our best QMC results (QMC/CCcbs and, when available, QMCcbs) are consistent with the newly available, and presumably of higher quality, experimental data.  While demonstrating consistency with past results is obviously a necessary phase in the development of any new method, we are certainly encouraged by the predictive capability already shown by our methodology.


We now proceed to highlight a number of notable cases:

de Oliveira-Filho and coworkers do not consider VH, presumably because the experimental dissociation energy has not been measured spectroscopically.  In Fig. \ref{fig:V} we show the experimental value selected by Truhlar and coworkers\cite{truhlar} which is derived from enthalpies of formation.  We prefer this number over those proposed in Ref. \citenum{fang2017prediction}, since the disparity of values shown in Ref. \citenum{fang2017prediction} (52.5 $\pm$ 4.4 and 56.1 $\pm$ 1.5 kcal/mol) illustrates the sensitivity of $D_e$ to the R-dependent quantity $BDE_{heterolytic}$(R-H), as discussed in Sec. IV.  

For CrO, even though our QMC/CCcbs result agrees with the value chosen by de Oliveira-Filho and co-workers, our QMC/MP2cbs and QMCcbs value deviate from that value by 4(2) and 5(2) kcal/mol, respectively.  However, we note two alternative experimental values:  111.1 $\pm$ 2 reported by Dixon and co-workers, and 110 $\pm$ 2 reported using ion molecule reactions in Ref. \citenum{kang1986gas}.  Both of these experiments are consistent with our best method, QMCcbs, which gives 110.1 $\pm$ 1.3 kcal/mol.

For CoS, single-reference CCSD(T), the ``gold standard'' to many, is in error by a sizable 14.6 kcal/mol.  The large MR correction, in the TZ basis, brought the CCSD(T) value within error bars of experiment.  ph-AFQMC calculations, which also show excellent agreement with experiment, strengthen our confidence in the reliability of the experimental measurement.

For CoO, CCSD(T) again makes a large error of 10.9 kcal/mol.  The MR correction, however, is not sufficient this time, as the MR-CCSD(T) result is still off by 3.6 kcal/mol.  de Oliveira-Filho and coworkers suggest that an MR correction in a larger basis may fix this.  We also note that the scalar relativistic correction is rather large for this molecule, at -4.5 kcal/mol.  All CBS extrapolation variants of our ph-AFQMC calculations produce values in agreement with the experimental value.  

For NiH, CCSD(T) is 8.5 kcal above experiment, and MR-CCSD(T) brings the BDE further away from experiment by an additional 3 kcal/mol.  Our QMC/MP2cbs result has a deviation comparable to that from MR-CCSD(T), and while QMC/CCcbs provides a substantial improvement, it is still off by 7.3 kcal/mol.  Only QMCcbs brings NiH within the error bars of experiment.  We note that our CS calculation in the QZ basis, with the CAS(11e,19o) trial, resolved two measurable plateaus.  Since it is not feasible to include more orbitals in the active space in this case, we measure the first plateau, as done (and justified) previously in Ref. \citenum{shee2018gpu}.       

CuO is another difficult case for all but our best full QMC treatment.  CCSD(T) is off by 9.3 kcal/mol, and MR-CCSD(T) by 8.4 kcal/mol. QMC/MP2cbs, QMC/CCcbs, and QMCcbs differ from experiment by 4, 5.5, and 1.6 kcal/mol.  

\subsection{Selection of Best Values and Comparisons of ph-AFQMC with DFT and CC methods}

With the goal of obtaining a robust, objective statistical comparison among ph-AFQMC, CCSD(T), MR-CCSD(T), and various DFT functionals, we now lay out a protocol to  construct a reliable test set of reference experimental values.

We begin with a baseline set of experiments selected by de Oliveira-Filho and co-workers\cite{aoto2017arrive}.  When possible, we substitute the high quality experimental results obtained via the predissociation technique of Ref. \citenum{morse2018predissociation}.  We then consider the independently selected best experimental values in the work of Dixon and co-workers\cite{fang2017prediction}, and remove from the test set any case for which there is disagreement, considering error bars, with the original experiment from either de Oliveira-Filho or Morse.  This situation arises only once, for CrO, and we therefore exclude it from our comparison. 

For ZnS, the BDEs computed with QMC/MP2cbs, QMC/CCcbs, and MR-CCSD(T) are all in excellent agreement, with a value roughly 15 kcal/mol below experiment, which is 49.1 $\pm$ 3 kcal/mol as suggested by de Oliveira-Filho and co-workers\cite{aoto2017arrive} and derived from the work of Marquat and Berkowitz \cite{marquart1963dissociation} and de Maria et al.\cite{de1965mass}. This experimental value is in stark contrast to the value of 34.3 $\pm$ 1.0 kcal/mol suggested by Truhlar and co-workers\cite{truhlar}. The latter value, as pointed out by de Oliveira-Filho and co-workers\cite{aoto2017arrive}, is derived from a thermochemical analysis by von Szentpaly\cite{von2008atom}, which used theoretical, not experimental, values provided by Peterson et al.\cite{peterson2007group}  Papakondylis, who used various theoretical methods to calculate the $D_e$ of ZnS,\cite{papakondylis2011ab} pointed out that the aforementioned experimental papers used outdated values for the equilibrium bond lengths and frequencies as compared to more recent measurements,\cite{hou2010studies,zack2009pure} bringing the older experimental measurements into doubt.  Therefore, more experimental investigation into ZnS should be done to see if indeed the calculations are correct.  For these reasons, we omit ZnS from our analysis, and simply report to the community our prediction of 38.8 $\pm$ 2.4 kcal/mol, as obtained with our QMCcbs method.  We believe that the quantitatively consistent results provided by completely independent high-level electronic structure approaches will eventually prove to predict the correct experimental value, but this speculation will have to wait for further experiments to confirm or refute.  

The MAE of all QMC methods, the CC approaches, and 10 DFT functionals, with respect to the experimental set of 40 molecules as selected above, are given in Table \ref{table:MAEs}.  

\begin{table}[H]
\begin{threeparttable}
\caption{Mean Absolute Error and Maximum Error on $D_e$ shown for AFQMC, CCSD(T), icMRCCSD(T), and DFT methods vs. the experiments selected in Ref. \citenum{aoto2017arrive} and, when possible, Ref. \citenum{morse2018predissociation}.  For reasons justified in the text, we omit VH, CrO, and ZnS from the comparative statistical analysis.  In all, our test set contains 41 diatomics.  All DFT calculations are in the aug-cc-pVQZ basis with DKH corrections.  DFT and CC values taken from Ref. \citenum{aoto2017arrive}.  All values are in kcal/mol.}. 
\begin{tabular}{l l r}
\hline\hline
                      Method &  MAE  & Max Error \\ [0.5ex] 
\hline
PBE    & 15.66 & 40.90 \\
BP86   & 14.78 & 38.17 \\
TPSS   & 12.83 & 31.00 \\
M06-2X & 12.05 & 37.95 \\
BLYP   & 11.64 & 37.10 \\
M06-L  & 8.44  & 21.85 \\
M06    & 7.06  & 22.25 \\
PBE0   & 4.73  & 21.85 \\
B3LYP  & 4.45  & 23.45 \\
B97    & 3.70  & 17.25 \\
CCSD(T)& 2.84  & 17.35 \\
icMRCCSD(T)       & 2.76    & 11.60 \\
QMC/MP2cbs        & 2.3(4)  & 12(2) \\
QMC/CCcbs         & 2.1(4)  & 7(2) \\
QMC/MP2+QZcbs\tnote{$\dagger$} & 1.5(4) & 4(3) \\
QMC/CC+QZcbs\tnote{$\dagger$}  & 1.4(4) & 3(3) \\
\hline
\end{tabular}
\begin{tablenotes}
\item[$\dagger$] Includes QMC TZ/QZ CBS extrapolations when available
\end{tablenotes}
\label{table:MAEs}
\end{threeparttable}
\end{table}

We note that the treatment of spin-orbit effects at the CASSCF level does not include dynamic correlation, which can in some cases be very large (e.g., -3 kcal/mol for NiCl, -3.1 for NiO, -2.4 for NiF).  For reasons such as this, DeYonkers and co-workers have suggested that ``chemical accuracy" for transition metal species is $\pm 3$ kcal/mol.\cite{deyonker2007quantitative}  None of the DFT functionals considered in this study meets this criterion.  As is well known,  DFT results are highly dependent on the exchange-correlation functional used, with MAEs ranging from 3.7 to 15.7 kcal/mol.  The highest level of accuracy is obtained with the B97 functional, and our data suggests that it should be chosen in DFT studies of similar transition metal chemistries.  

The results of this work would argue that Truhlar's original claim, that CCSD(T) and DFT produce comparable accuracy, must be qualified.  A head to head comparison of the B97 functional in a large basis set with state-of-the-art CCSD(T) in the CBS limit shows that while both exhibit equally large maximum errors ($\sim$17 kcal/mol), the MAE of the latter is slightly, but significantly, lower.  That said, the accuracy of DFT depends entirely, and perhaps unsystematically, on the functional employed, and we note that in comparison to the majority of functionals, CCSD(T) should be preferred assuming one has adequate computing capacity. 

For the CCSD(T) approach and its MR-corrected variant, our analysis gives MAEs of 2.84 and 2.76 kcal/mol, and maximum errors of 17.35 and 11.6 kcal/mol, respectively.  A robust benchmark method for transition metal chemistry, in our view, cannot make such large errors for individual cases.  That is \emph{not} to say that CCSD(T) is necessarily unfit for benchmark applications.  Specifically, both the CCSD(T) protocol and the MR-CCSD(T) results we present, as performed in Ref. \citenum{aoto2017arrive}, involve a number of assumptions that may lead to suboptimal accuracy.  Chief among them are the additivity assumptions involving the core-valence and scalar relativistic corrections.  In the CC protocol, the CBS limit is estimated with the non-relativistic Hamiltonian.  Then, a relativistic correction using the DKH Hamiltonian in a TZ-level basis is added.  Similarly, the MR-CCSD(T) values shown simply add a multi-reference corrections computed in the TZ basis without core-valence effects treated explicitly, and without the DKH Hamiltonian and corresponding basis sets.  Indeed, the relativistic corrections can be quite large, e.g. -8.1, 6.4, and -7.8 kcal/mol for NiCl, CoCl, and NiF, respectively.  In such cases, among others, it is plausible that the additivity assumptions mentioned above break down.  

Another potential source of error is that the MR-CCSD(T) calculations have not been converged with respect to active space size, likely due to the high computational expense associated with such a procedure.  Indeed, the full-valence (and sometimes smaller) active spaces may be insufficient for cases in which excitations into high-lying virtual and/or from low-lying occupied orbitals contribute significantly to the correlation energy.  

Thus, these points suggest that more accurate CC results are possible, in principle, but only if one is willing to bear the high computational expense required to carry out a more rigorous computational protocol.

Turning to our ph-AFQMC methods, we first notice that QMC/MP2cbs, with an MAE of 2.3(4) kcal/mol and maximum error of 12(2) kcal/mol, is of comparable quality to, or arguably slightly more robust than, MR-CCSD(T).  This is remarkable given that the latter involves CC calculations, which scale as the seventh power with system size, in QZ and 5Z basis sets, while the former requires a ph-AFQMC calculation in the TZ basis only, followed by a relatively inexpensive two-point MP2 extrapolation.  The near-perfect parallel efficiency of the QMC calculation, and its acceleration on graphical processing units, are advantages that are not enjoyed by traditional CC implementations. 

QMC/CCcbs achieves notable reductions in both the MAE and maximum error, at 2.1(4) and 7(2) kcal/mol, respectively.  We note that for larger systems, using localized orbital implementations of CCSD(T) can drastically reduce the computational cost, while preserving systematic improvability with regard to localization errors.  For systems with substantial MR character, methods such as CASSCF of selected CI supplemented with perturbation theory can be used to replace CCSD(T) to perform the CBS extrapolation.

When the 13 cases for which we performed CBS extrapolations entirely with ph-AFQMC are taken into account, the maximum error of our best method, QMC/CC+QZcbs is reduced to 3(3) kcal/mol, with an MAE of 1.4(4) kcal/mol.  For larger systems, the QZ extrapolation option becomes relatively more advantageous, given its accuracy at low-polynomial scaling.  

\section{Conclusions and Outlook}

In summary, we have computed the $D_e$ of 44 3$d$ transition metal-containing diatomic molecules with ph-AFQMC.  We describe the extension of a recently developed correlated sampling approach to the calculation of bond dissociation energies, and report improvements in both efficiency and accuracy compared to uncorrelated calculations.  In order to assess the robustness of various CBS extrapolation techniques, and moreover to compare our ph-AFQMC results to the DFT, CCSD(T), and MR-CCSD(T) calculations performed in Ref. \citenum{aoto2017arrive}, we carefully assemble a set of reference experimental values via the following unbiased protocol.  We use the experimental values selected by de Oliveira-Filho and co-workers, and, when available, predissociation measurements from a recently published work by Morse and co-workers.  VH was omitted in Ref. \cite{aoto2017arrive}, thereby depriving us not only of a consistently-chosen experimental reference but also of consistently-computed DFT and CC values, so it is not included in our statistical analysis.  We omit cases where the experimental values selected by de Oliveira-Filho and Dixon are significantly different (with non-overlaping error bars), which necessitates the removal of CrO from the test set.  Finally, we remove ZnS from the analysis on the grounds of concerns regarding the validity of the reported experimental number, which has been voiced previously in the literature, and emboldened substantially by the observation that ph-AFQMC, CCSD(T), and MR-CCSD(T) all disagree with the experimental value, and roughly agree with each other given statistical error bars.  

Using this set of reference values, we assess the accuracy of our ph-AFQMC calculations alongside previously published results from 10 DFT functionals, CCSD(T), and MR-CCSD(T).  We find that of the DFT functionals, B97 performs best, and suggest its use for future DFT studies of transition metal-containing systems.  We find that CC methods, while more accurate on average than DFT approaches, are not suitable benchmark methods for these systems due to the persistence of outliers with errors in excess of 10 kcal/mol.  We take advantage of the systematic improvability of the ph-AFQMC method to attain high-quality predictions for these diatomic systems, and experiment with various cost-effective CBS extrapolation methods utilizing MP2 or CC.  The final MAE of our best calculations is 1.4(4) kcal/mol, with maximum error of 3(3) kcal/mol.  We would like to draw particular attention to the need for more robust experimental determinations of the dissocation energy of ZnS (as discussed before) and CrO, as, with regard to the latter, our most reliable ph-AFQMC calculation predicts 110.1 $\pm$ 1.3 kcal/mol, which is in agreement with two other published experiments,\cite{fang2017prediction,kang1986gas} but not the one put forth by de Oliveira-Filho and co-workers.  

Although they are composed only of two atoms, these transition metal systems exhibit very complex electronic structures, with a wide range of both static and dynamic correlation, core-valence and relativistic phenomena.  The presence of many competing, low-lying states is common in transition metal-containing systems, e.g. in the Ni atom\cite{botch1981valence} and FeS,\cite{scemama2018deterministic} and we have shown that our ph-AFQMC protocol is capable of constraining calculations to targeted, experimentally-observed angular momenta and spatial symmetries that characterize the ground states.   

That our QMC calculations can achieve such high accuracy is even more remarkable given that the trial wavefunctions used to implement the phaseless constraint utilize between 100 and 5700 determinants ($\sim$800 on average).  However, obtaining CASSCF wavefunctions with sufficiently many active electrons and orbitals will become a challenge when larger systems are considered.  We stress the need to experiment with alternative trial wavefunctions, the choice of which will likely depend on the target application.  We are optimistic that explorations into more efficient descriptions of dynamic correlation and ways to exploit the locality of entanglements will lead the way toward scalable trials for accurate ph-AFQMC calculations.

\begin{acknowledgement}
JS gratefully acknowledges Mario Motta and Hao Shi for helpful insights. D.R.R. acknowledges funding from NSF CHE-1839464.
S.Z. acknowledges funding from DOE DE-SC0001303. 
This research used resources of the Center for Functional Nanomaterials, which is a U.S. DOE Office of Science Facility, and the Scientific Data and Computing Center, a component of the Computational Science Initiative, at Brookhaven National Laboratory under Contract No. DE-SC0012704 
This research used resources of the Oak Ridge Leadership Computing Facility, which is a DOE Office of Science User Facility supported under Contract DE-AC05-00OR22725.  The Flatiron Institute is a division of the Simons Foundation. 
\end{acknowledgement}

\begin{suppinfo}


\end{suppinfo}

\bibliography{References}

\end{document}